\documentclass[twocolumn,showpacs]{revtex4}      
\usepackage{wrapfig,epsfig,epsf}
\usepackage{dcolumn}
\usepackage{amssymb,amsmath}
\usepackage{mathtext}
\usepackage{rotate}
\usepackage{euscript}
\usepackage{bm}

\begin{document}

\input epsf
\renewcommand{\topfraction}{0.8}

\title {\Large\bf On some constraints on inflation models with power-law potentials }
 \author{\bf S. A. Pavluchenko}
\affiliation{ { Sternberg Astronomical Institute, Moscow State University, Moscow 119992, Russia}    }

{\begin{abstract}
We investigate inflation in closed Friedmann-Robertson-Walker universe filled with the scalar field
with power-law potential. For a wide range of powers and parameters of the potential we numerically
calculate the slow-roll parameters and scalar spectral index at the epoch when present Hubble scale
leaves the horizon and at the end of inflation. Also we compare results of our numerical calculations 
with recent observation data. This allows us to set a constraint on the power of the potential: 
$\alpha \lesssim (3.5 \div 4.5)$. 
\end{abstract}}

\maketitle

\section{Introduction}

Inflation, the stage of accelerated expansion of the universe, first proposed in the begin of 
1980th~(\cite{infl_first}), nowadays attempts a great attention as well. In our previous 
paper~\cite{my4} we investigated the generality of inflation for a wide class of quintessence 
potentials and noted that criteria we used 
is weak to decide about the inflationarity of the model with given potential. So in this 
paper we are proceeding with the investigation of closed Friedmann-Robertson-Walker (FRW) models, but 
using power-law potentials only. 
In this way we calculate spectral indices, slow-roll parameters and other values on the epoch, 
when the present Hubble scale leaves the horizon and at the end of inflation, to compare our
constraints with early obtained results and with observation data.
This allows us to make some constraints on FRW models with 
scalar field, on potentials of the scalar field and on parameters of these potentials. Since due to
inflation universe becomes flat very quickly our constraints are applicable to flat case as well.
If one 
supposes that the same potential describes both the inflation stage in early universe and acceleration
nowadays we can compare our constraints with constraints on the quintessence potentials (see 
e.g.~\cite{Liddle1,als1} or \cite{revs1,revs2,revs3} for review of the problem).

Another aim of this paper is the investigation of the influence of the initial conditions on inflation 
dynamics in case of closed FRW models. Really, in case of initially flat universe we have
only one degree of freedom~-- how does primordial energy density distribute between kinetic and 
potential terms, but in case of initially closed universe we have at least one more degree of
freedom~-- the distribution between curvature and initial expansion rate terms. So our second aim is in
studying these two distributions, comparison them to each other and investigation how does they 
act on inflation.

\section{Main equations}

The equations describing the evolution of the universe in closed FRW model are
$$
\frac{m_P^2}{16 \pi}\left(\ddot a+\frac{{\dot a}^2}{2a}+\frac{1}{2a} \right) + \frac{a}{4} \left( 
\frac{{\dot \varphi}^2}{2}- V(\varphi) \right) =0,  
$$
$$
\ddot \varphi + \frac{3 \dot \varphi \dot {\vphantom{\varphi}a} }{a}+\frac{dV(\varphi)}{d\varphi} =0, 
$$
\noindent and the first integral of the system is
$$
\frac{3 m_P^2}{8 \pi}\left(\frac{\dot a^2}{a^2} + \frac{1}{a^2}\right)=\left( V(\varphi)+\frac{{\dot 
\varphi}^2}{2} \right), 
$$

\noindent where $m_P = 1/\sqrt{G} = 1.2 \times 10^{19} {\rm GeV}$.

And we use the {\it trigonometrical (angular)} parameterization $(\phi,H_0)$ of the space of initial 
conditions:
$$
\frac{3 m_P^2}{8 \pi} \left( H_0^2 + \frac{1}{a^2} \right) = m_P^4, \quad 
H_0^2 + \frac{1}{a^2} = \frac{8 \pi m_P^2}{3}, \eqno(1)
$$
$$
H_0 \in \left[ 0; \sqrt{\frac{8\pi}{3}}m_P \right); 
$$
$$
V(\varphi) + \frac{ {\dot \varphi}^2 }{2} = m_P^4, \quad V(\varphi) = m_P^4 \cos^2 \phi, 
$$
$$
\frac{ {\dot \varphi}^2 }{2} = m_P^4 \sin^2 \phi, \quad \phi \in \left[ -\frac{\pi}{2}; \frac{\pi}{2}
\right]. 
$$

Our method is as follows. Like in~\cite{my4} we start from the Planck boundary for a given pair of 
initial conditions $(\phi, H_0)$ and then numerically calculate the further evolution of the universe
through inflation. Also we calculate scalar spectral index and slow-roll parameters 
during the evolution of the universe.
And since the universe
becomes flat very quickly due to the inflation, we can use the usual determination for scalar spectral 
index.
There are two expressions for it~-- first order~\cite{L&L1} and
second order~\cite{S&L1} ones and we use second order results for more precise calculations. 
Also one can express them using {\it Potential Slow-Roll Approximation}
(PSRA)~\cite{ST,L&L1} and {\it Hubble Slow-Roll Approximation} (HSRA)~\cite{HSRA} (see~\cite{Barrow1}
for details). Equations for slow-roll parameters are:

\begin{figure*}
\epsfxsize=18cm
\centerline{\epsfbox{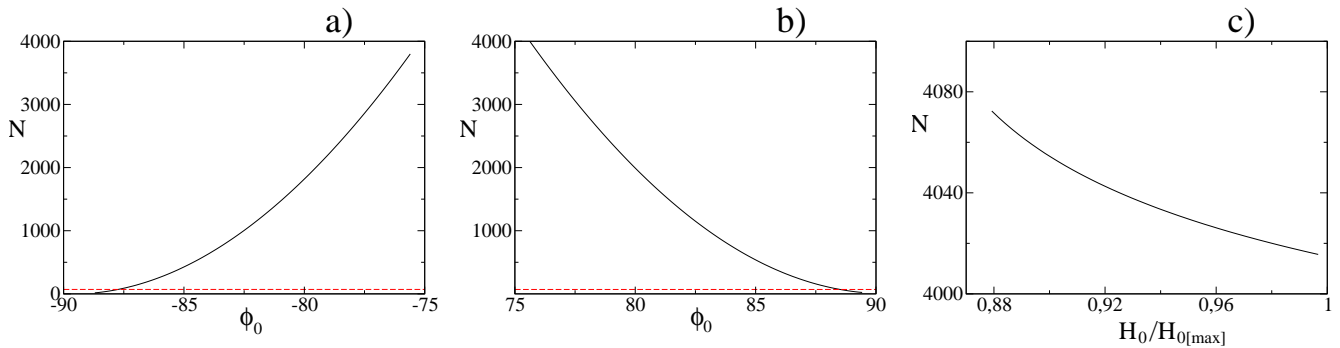}}
\caption
{The dependence of the number of e-foldings universe experienced during inflation on the initial
conditions for the angular parameterization. It's clear that the dependence on $\phi$ ((a) for the 
case of positive $\phi$ and (b) for the case of negative $\phi$) is stronger that the dependence on $H$ (c)
(see text for details).
}
\end{figure*}

\noindent for PSRA
$$
\epsilon_V(\varphi) = \frac{m_P^2}{16\pi} \left( \frac{V'(\varphi)}{V(\varphi)}\right)^2,
$$
$$
\eta_V(\varphi) = \frac{m_P^2}{8\pi} \frac{V''(\varphi)}{V(\varphi)};
$$
\noindent for HSRA
$$
\epsilon_H(\varphi) = \frac{m_P^2}{4\pi} \left( \frac{H'(\varphi)}{H(\varphi)} \right)^2,
$$
$$
\eta_H(\varphi) = \frac{m_P^2}{4\pi} \frac{H''(\varphi)}{H(\varphi)};
$$

\noindent and for $n_s$ respectively:

\begin{figure*}
\epsfxsize=18cm
\centerline{\epsfbox{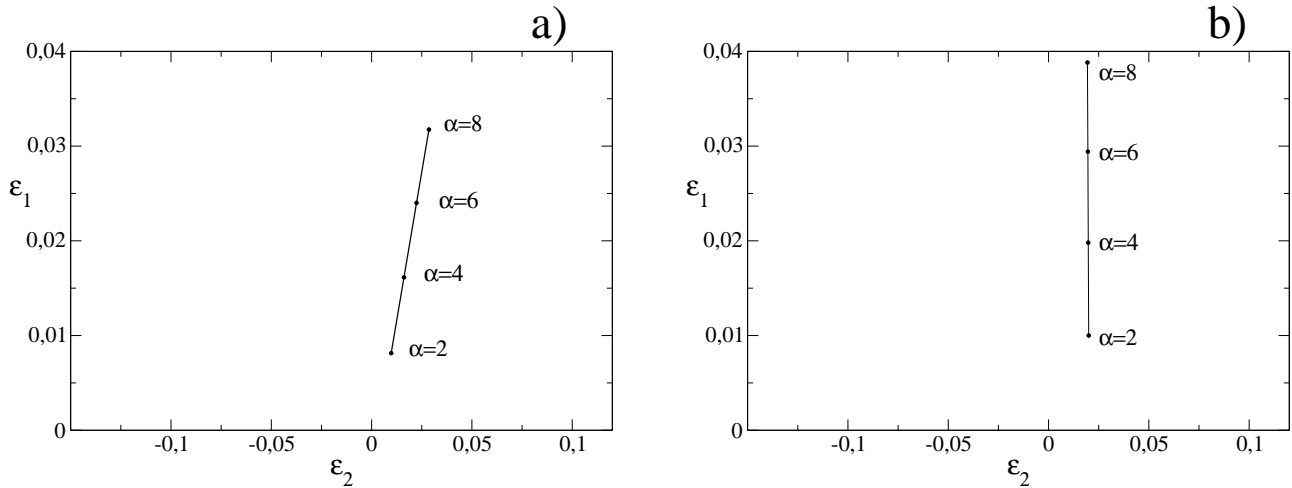}}
\caption
{Positions of models with different powers of power-law potentials on ($\epsilon_1, \epsilon_2$) plane
in case of $N_{\rm hor} = 62$ at (a) panel and $N_{\rm hor} = 50$ at (b) panel.
}
\end{figure*}

\begin{multline*}
n_s = 1 - 6\epsilon_V + 2\eta_V + \frac{1}{3}(44-18c)\epsilon_V^2 +\\ +(4c-14)\epsilon_V\eta_V + 
\frac{2}{3}\eta_V^2 + ... ,   \tag{2}
\end{multline*}
\begin{multline*}
n_s = 1 - 4\epsilon_H + 2\eta_H - 2(1+c)\epsilon_H^2 - \\  -\frac{1}{2}(3-5c)\epsilon_H \eta_H 
-\frac{1}{2}(3-c)\eta_H^2 + ... ,   \tag{3}
\end{multline*}

\noindent where $c \equiv 4(\ln 2 + b) -5 \simeq 0.08145$ with Euler-Mascheroni constant $b$. So
we calculate $n_s$ by both these cases and compare them to each other. Also we calculate
slow-roll parameters and compare them at the epoch when the present Hubble scale leaves the horizon 
for different potentials and for different parameters of the potentials.

Since the purpose of our paper is to set constraints on inflation models, we compare
results of our numerical calculations with constraints on $n_s$ and other values from experiments.
In particular we compare them with results
obtained in Ref.~\cite{Liddle1}, so we need to introduce horizon-flow parameters $\epsilon_1$, 
$\epsilon_2$ (see~\cite{Liddle_add1,Liddle_add2} for more details):
$$
\epsilon_1 \simeq \frac{m_P^2}{16\pi} \left( \frac{V'}{V} \right)^2, \eqno(4)
$$
$$
\epsilon_2 \simeq \frac{m_P^2}{4\pi} \left[ \left( \frac{V'}{V} \right)^2 - \frac{V''}{V}  \right].
\eqno(5)
$$

To compare our constraints with~\cite{Liddle1} we need to localize the moment during 
inflation when the present Hubble scale leaves the horizon. It took place at about 60 e-folds
before the end of inflation~(\cite{efolds}, see also~\cite{L&L2} for details). This value is 
model-dependent and it depends also on the way of inflation ends so to simplify we use
two values for this~-- 62 as a bound value and 50. And so we calculate all parameters on these
two epochs~-- 62 and 50 e-folds before the end of inflation. For further using we denote this 
value as $N_{\rm hor}$.

\section{Power-law potential}

First, let us describe the dependence of the number of e-foldings universe experienced during inflation
on initial conditions. To illustrate this
we plotted in Fig.~1 this dependence on $\phi$ and on $H$ separately; to realize the whole
picture one needs to multiply these functions. So in Fig.~1(a) there is a dependence of the number
of e-foldings universe experienced during inflation on $\phi$ for negative $\phi$, in Fig.~1(b) 
the same but for positive $\phi$ and in Fig.~1(c)~-- on $H_0$. From Fig.~1(a) and (b) one can also 
see the influence of sign of initial $\dot \varphi$~-- positive $\phi$ corresponds to positive initial
$\dot \varphi$ and negative $\phi$ corresponds to negative initial $\dot \varphi$. For instance, for
$H_0 = H_{0[max]}$~-- the flat case~-- measure of trajectories experienced insufficient inflation (the case 
when universe experienced inflation but the number of e-foldings is less then 70) is about 
$2.26^\circ$ 
for negative $\phi$ and about $1.49^\circ$ for positive $\phi$ (so for flat case the universe 
experienced
sufficient inflation for $-87.75^\circ < \phi <88.51^\circ$; in Fig.~1(a) and 1(b) dashed line  
corresponds namely to $N=70$ e-foldings).

From Fig.~1 one can also learn that the dependence of the number of e-foldings on $H_0$ is many times 
weaker than the dependence on $\phi$. Really, $\phi$ determines initial distribution of the energy
between kinetic and potential terms and, since the universe very quickly reaches the slow-roll regime,
it determines the energy density at the beginning of inflation. Also during the reaching slow-roll 
regime 
$H$ becomes large, so $H_0$~-- initial value of $H$ weakly acts on the energy density at the beginning 
of inflation.


\begin{figure*}
\epsfxsize=18cm
\centerline{\epsfbox{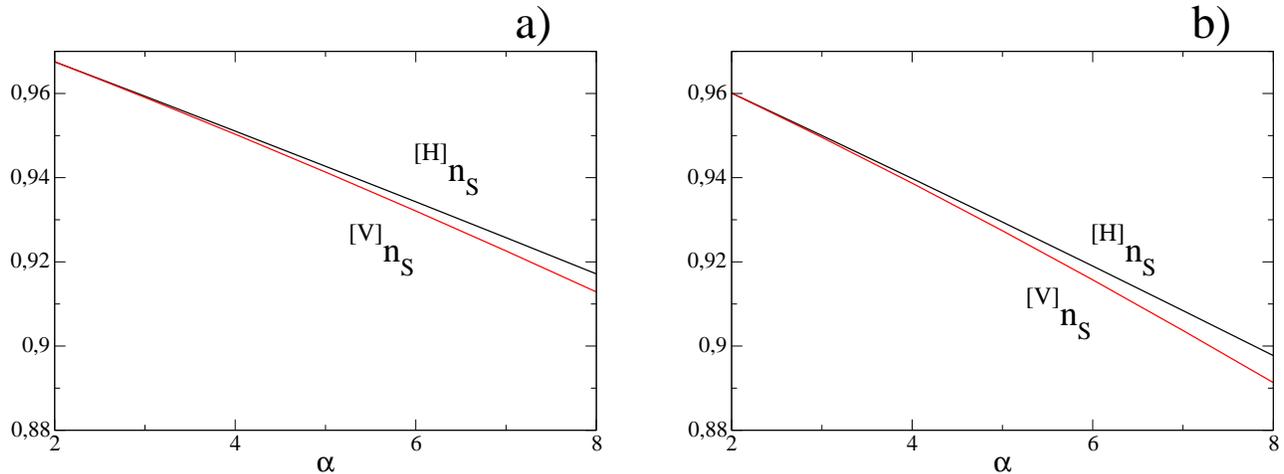}}
\caption
{The dependence of $n_s$, calculated in PSRA and HSRA (see text for details), on power $\alpha$ in
case of $N_{\rm hor}=62$ at (a) panel and in case of $N_{\rm hor}=50$ at (b) panel. 
}
\end{figure*}

Power-law potentials are well studied and they lead to "chaotic inflation"~\cite{chaotic}. One can 
really use them as inflation part in potentials like those considered by Peebles and 
Vilenkin~\cite{peeb_vil_99}. They have also attracted attention for some of their 
properties~\cite{class,Kolda_Lyth99}.

We consider power-law potentials of a kind
$$
V(\varphi) = \frac{\lambda \varphi^\alpha}{\alpha} = \lambda^* \left( \frac{\varphi}{m_P} 
\right)^\alpha, \eqno(6)
$$

\noindent and our results are as follows. In Fig.~2 we plotted positions of the models with
different powers on ($\epsilon_1, \epsilon_2$) plane in case of $N_{\rm hor}=62$ at (a) panel and
in case of $N_{\rm hor}=50$ at (b) panel. By comparison these plots with bounds on
($\epsilon_1, \epsilon_2$) plane obtained
in~\cite{Liddle1} from 2dF and WMAP data one can make a constraint on $\alpha$ as $\alpha 
\lesssim 4.8$ in case of
$N_{\rm hor} = 62$ and $\alpha \lesssim 3.5$ in case of $N_{\rm hor} = 50$.

Another constraint on power $\alpha$ one can obtain from Fig.~3. In Fig.~3 we plotted the 
dependence of $n_s$: calculated in PSRA (see eq.~(2)) we denoted as $^{[V]}n_s$ and calculated 
in HSRA (see eq.~(3)) we denoted as $^{[H]}n_s$. Using bounds on $n_s$~\cite{WMAP}:
$n_s = 0.99 \pm 0.04$ (WMAP only) and $n_s = 0.97 \pm 0.03$ (WMAP+ACBAR+CBI+2dFGRS+$\rm 
L_\alpha$-forest) one can set a bound $\alpha \lesssim 4.5$ in case of $N_{\rm hor} = 62$ and
$\alpha \lesssim 3.8$ in case of $N_{\rm hor} = 50$. One can see these two constraints~-- from 
$n_s$ and previous one~-- are close to each other.

Now let us set a constraint on $\lambda^*$. To do this we can use results obtained from COBE data
in~\cite{Liddle_add2}:                                                
$$
3 < \frac{V^{1/4}_{\rm infl}}{10^{15}{\rm GeV}} < 29, \eqno(7)
$$

\noindent and we calculate these values at the end of inflation. 
After defining $K(\alpha) = (\varphi_{\rm end}/m_P)$ one can obtain $V_{\rm end} = \lambda^* 
K^\alpha(\alpha)$ and after substitution (6) to (7):
$$
\lambda_2^* < \lambda^* < \lambda_1^*,
$$
\noindent where
$$
\lambda_1^* = 3.4 \times 10^{-11} m_P^4 K^{-\alpha}(\alpha), 
$$
$$
\lambda_2^* = 3.9 \times 10^{-15} m_P^4 K^{-\alpha}(\alpha).
$$

One needs to keep in mind the relation between $\lambda$ and $\lambda^*$ to recalculate 
$\lambda^*$ into $\lambda$ and inversely:
$$
\lambda = \frac{\alpha \lambda^*}{m_P^\alpha}, \quad \lambda^* = \frac{\lambda m_P^\alpha}{\alpha}.
$$


Another test, also linked with COBE normalization~\cite{COBE}, is about density perturbation spectrum 
$A_S(k)$:
$$
A_S^2(k) = \left. \frac{512\pi}{75} \frac{V^3}{m_P^6 {V'}^2} \right|_{k=aH}, \eqno(8)
$$

\noindent and this value is calculated on the epoch when the present Hubble scale leaves the horizon.
Also we can define value of the field on this epoch as $\varphi_{\rm hor} = L(\alpha) m_P$ 
and so rewrite (8) using (6) as
$$
A_S^2 = \frac{512\pi}{75 m_P^4} \frac{\lambda^* L^{\alpha+2}(\alpha)}{\alpha^2}.
$$
Let us remind the reader that according to COBE data this value is about $A_S \approx 2 \times 
10^{-5}$. Using it one can get another estimation for $\lambda^*$:
$$
\lambda^*_{A_S} = \frac{3 \times 10^{-8} m_P^4 \alpha^2}{512\pi L^{\alpha+2}(\alpha)}.
$$

\begin{figure}
\epsfxsize=7.5cm
\centerline{\epsfbox{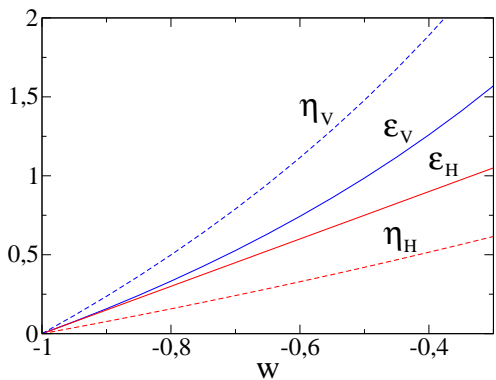}}
\caption
{An example of the behavior of slow-roll parameters with respect to 
the equation of state of the scalar field near
the end of inflation. It's clear that at the moment when inflation ends ($w=-1/3$) only $\epsilon_H$
is equal to unity but other parameters do not (see text for details).
}
\end{figure}

\begin{figure*}
\epsfxsize=12.6cm
\centerline{\epsfbox{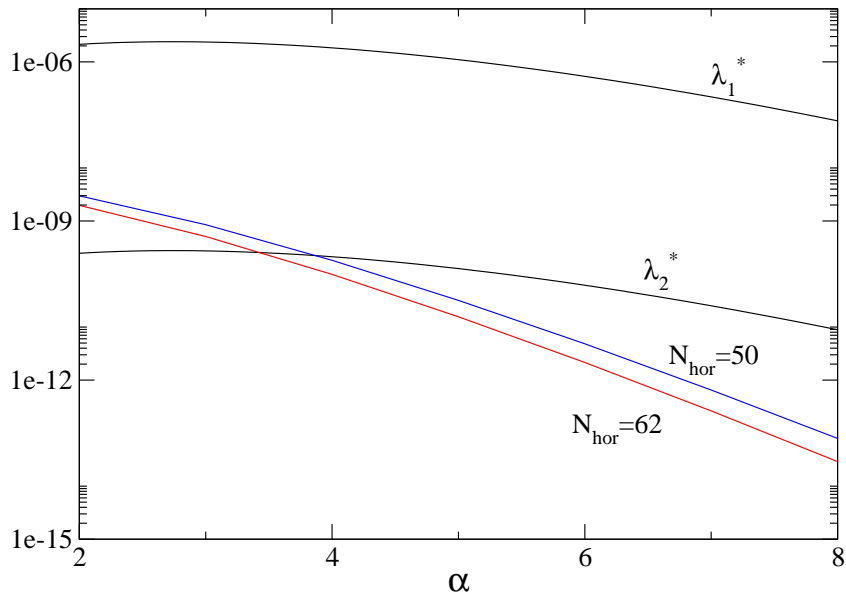}}
\caption
{The dependence of $\lambda_1^*$ and $\lambda_2^*$ as well as $\lambda^*_{A_S}$,
calculated for $N_{\rm hor} = 62$ and $N_{\rm hor} = 50$, on the power $\alpha$ (see text for details).
}
\end{figure*}

Finally, in Fig.~5 we plotted the dependence of both $\lambda_1^*$ and $\lambda_2^*$ on $\alpha$ 
and the dependence of $\lambda^*_{A_S}$ on $\alpha$. The range of possible values of $\lambda^*$ due
to eq.~(7)(\cite{Liddle_add2})
is between curves $\lambda_1^*$ and $\lambda_2^*$.
We plotted $\lambda^*_{A_S}$ for cases
$N_{\rm hor} = 62$ and $N_{\rm hor} = 50$. One can see from Fig.~5 that in case of $N_{\rm hor}
= 62$ we have a constraint $\alpha \lesssim 3.5$ and in case of $N_{\rm hor} = 50$ we have a
constraint $\alpha \lesssim 4.0$.

\section{Conclusions}

So we reached our aim~-- we set some constraints on power-law potentials and their parameters.
We calculated the whole evolution of the universe during inflation for a wide range of 
initial conditions, parameters of power-law potentials (power $\alpha$ and $\lambda$) and set some 
constraints
on power-law potentials and on their parameters. Also we compared our constraints with results
obtained from the recent cosmic microwave background (CMB) data and large scale structure (LSS) 
data~\cite{Liddle1,L&L2,efolds}.
As we noted above for an epoch when the present Hubble scale leaves the horizon we have used two
values~-- 62 e-folds before inflation ends and 50 e-folds. And 62 e-folds is bound value in the sense
that other possible values are smaller then 62. We used 50 e-folds namely as an example of such a 
value and to demonstrate what can happen with $n_s$, $\epsilon_1$ and other values if we use lower 
(then 62) number of e-folds before the end of inflation to determine the epoch when the present Hubble scale 
leaves the horizon. And the result we obtained is: $\alpha \lesssim (3.5 \div 4.5)$. The exact value is
very model dependent. It depends on many factors, first on the way of inflation ends~-- this 
determines the
epoch when the present Hubble scale leaves the horizon. From figures one can see how does it act on
the results. Also it depends on the observation data~-- to make our constraints more precise we need 
more
precise observation data. But even with these uncertainties our constraints are some harder then 
results obtained from the CMB and LSS data~\cite{Liddle1,L&L2,efolds}.

One can see that our numerical results are some differ from analytical results. This is due to the fact 
that one can incorrectly determine exact moment when inflation ends using relation $\rm max\{\epsilon,
\eta\} = 1$. As one can see from Fig.~4 (we plotted it only as an example; it corresponds to the case 
$\alpha = 4$) at {\it real} moment of the end of inflation only $\epsilon_H$ is exact equal to
unity. And since most analytical results are obtained using relation $\epsilon_V = 1$, it 
corresponds not to the exact moment of the end of inflation. And the small difference between 
our numerical results and analytical results is namely due to this uncertainty.

This work was supported by the Russian Ministry of Industry,
Science and Technology through the Leading Scientific School Grant $\#$ 2338.2003.2.
We would like to thank A.R. Liddle for useful and stimulating discussion
and N.Yu. Savchenko for useful discussion and different help in preparing this paper.

\end{document}